\documentclass[12pt]{article}
\usepackage{graphicx,color,epsfig,rotate}
\title{Phase transitions induced by correlated hopping in the
Falicov-Kimball model }
\author{Hana \v{C}en\v{c}arikov\'a, Pavol Farka\v sovsk\'y\\
Institute  of  Experimental  Physics,  Slovak   Academy   of
Sciences\\
Watsonova 47, 043 53 Ko\v {s}ice, Slovakia}
\date{}
\begin{document}
\baselineskip=20pt
\maketitle

\begin{abstract}
The extrapolation of finite-cluster calculations is used to examine ground-state 
properties of the one-dimensional Falicov-Kimball model with correlated hopping. 
It is shown that the correlated hopping strongly influences both the valence 
transitions and the conducting properties of the model and so it should not be 
neglected in the correct description of materials with correlated electrons.
This is illustrated for two selected values of the Coulomb interaction
that represent typical behavior of the model for small and intermediate
(strong) interactions. In both cases the insulator-metal transitions
(accompanied by continuous or discontinuous valence transitions) induced by 
correlated hopping are observed. 
\end{abstract}
\thanks{PACS nrs.:75.10.Lp, 71.27.+a, 71.28.+d, 71.30.+h}
\newpage

\section{Introduction}
Recent theoretical studies of the Falicov-Kimball model (FKM)~\cite{Falicov} 
showed that although this model is relatively simple, it can yield 
the correct physics for describing rare-earth and transition metal
compounds~\cite{Farky1, Theory}. The model is based on the coexistence of two
different types of electronic states in  given materials:
localized, highly correlated ionic-like states and extended,
uncorrelated, Bloch-like states. The Hamiltonian of the model
is given by sum of three terms 
\begin{equation}
H=\sum_{\langle ij\rangle}t_{ij}d^{+}_{i}d_{j} + U\sum_{i}d^{+}_{i}d_{i}f^{+}_{i}f_{i} +
E_f\sum_{i}f^{+}_{i}f_i.
\end{equation}

The first term of (1) is the kinetic energy corresponding to
quantum mechanical hopping of the itinerant $d$ electrons
between sites $i$ and $j$. Usually it is assumed that
$t_{ij}=-t$  if $i$ and $j$ are nearest neighbors and
zero otherwise (the conventional FKM), however, in what follows 
we consider a much more realistic type of hopping, 
so for the moment we leave it as arbitrary.
The second term represents the Coulomb interaction between $f$
electrons with density $n_f=\frac{N_f}{L}=
\frac{1}{L}\sum_{i}f^{+}_{i}f_{i}$ and $d$
electrons with density $n_d=\frac{N_d}{L}=
\frac{1}{L}\sum_{i}d^{+}_{i}d_{i}$, where
$L$ is the number of lattice sites.
The last term stands for the localized $f$ electrons whose sharp
energy level is $E_f$.

One can see, that the interaction term in the conventional FKM includes only 
the local interactions and all nonlocal interactions are neglected.
It is interesting to ask, however, if these nonlocal 
interactions are really unimportant and can be neglected or not. The
possible improvement of the conventional FKM, which  take into 
account also nonlocal interactions between nearest neighbors, is 
so-called correlated hopping term that describes how the occupancy 
of $f$ orbitals influences the hopping probability of $d$ electrons:
\begin{equation}
\tilde{t}_{ij}=t_{ij} + t'_{ij}(f^+_if_i + f^+_jf_j).
\end{equation}
The importance of the correlated hopping term has already been mentioned by
Hubbard~\cite{Hubbard}. Later Hirsch~\cite{Hirsch} pointed out that this 
term may be relevant in the explanation of superconducting properties of 
strongly correlated electrons. The effect of correlated hopping on the 
ground-state properties of the FKM has been examined 
recently by Gajek and Lema\'nski in 1D $\cite {GajLem}$ as well as Wojtkiewicz and Lema\'nski 
in two dimensions $\cite {WojtLem}$. 
The same subject has been studied in our previous papers~$\cite{FarkyGal,FarkyHud}$. 
Using finite-cluster diagonalization
calculations we have showed that the picture of valence and 
metal-insulator transitions is dramatically changed if the term of correlated 
hopping is included. One of the most important results found for the 
one-dimensional FKM with correlated hopping was that the correlated hopping 
can induce the insulator-metal transition, even in the half-filled band
case $n_d=n_f=1/2$~$\cite {FarkyHud}$, where the ground state of the conventional 
FKM is insulating for all Coulomb interactions $\cite {KenLieb}$. In the present 
paper we go beyond this case and examine a more general point $n_d+n_f=1$, 
that is the point of the special interest for valence and metal-insulator 
transitions caused by promotion of electrons from localized $f$ orbitals 
$(f^n \to f^{n-1})$ to the conduction band states.

Since in the spinless version of the FKM (1) with correlated hopping (2)
the $f$~-~electron occupation number $f^+_if_i$ of each 
site $i$ commutes with the Hamiltonian, the $f$-electron occupation 
number is a good quantum number, taking only two values: $w_i=1$ or 0, 
according to whether or not the site $i$ is occupied by the localized 
$f$~electron. Therefore the Hamiltonian of the FKM with correlated hopping can 
be rewritten as

\begin{equation}
H=\sum_{\langle ij\rangle}h_{ij}(w)d^+_id_j+E_f\sum_iw_i,
\end{equation}
where $h_{ij}(w)=\tilde{t}_{ij}(w)+Uw_i\delta_{ij}$ and

\begin{equation}
\tilde{t}_{ij}(w)=t_{ij}+t'_{ij}(w_i+w_j).
\end{equation}

Thus for a given $f$-electron configuration
$w=\{w_1,w_2, \dots ,w_L\}$ the Hamiltonian (3)
is the second-quantized version of the single-particle
Hamiltonian $h(w)$, so the investigation of
the model (3) is reduced to the investigation of the
spectrum of $h$ for different configurations of $f$ electrons.
This can be performed exactly (over the full set of $f$-electron 
configurations) or approximatively (over the incomplete set, but still
kipping the high precision of calculations). Here we adopt the second 
method since it allows us to treat several times larger lattices 
and so to minimize finite-size effects.
\section{The method}
The method used in this paper to study effects of correlated hopping 
on ground-state properties of the FKM is a simple modification of the
finite-cluster exact-diagonalization method. This method was
used firstly by one of us to describe the ground-state phase diagram 
(in $E_f$ - $U$ plane) of the conventional FKM in one and two 
dimensions $\cite {Farky2}$. It was shown that the method is able to reproduce 
satisfactorily  the exact results even after a relatively small number of 
iterations for both, one and two dimensions.
The method is relatively simple and consists of the following steps:
(i) Chose a trial configuration $w=\{w_1,w_2, \dots , w_L\}$.
(ii) Having $w$, $U$, $t'$ and $E_f$ fixed, find
all eigenvalues $\lambda_k$ of $h(w)=T+UW$. (iii) For a given
$N_f=\sum_iw_i$ determine the ground-state energy
$E(w)=\sum_{k=1}^{L-N_f}\lambda_k+E_fN_f$ of a particular
$f$-electron configuration $w$ by filling in the lowest
$N_d=L-N_f$ one-electron levels.
(iv) Generate a new configuration $w'$ by moving a randomly
chosen electron to a new position which is chosen also at random.
(v) Calculate the ground-state energy $E(w')$. If $E(w')<E(w)$
the new configuration is accepted, otherwise $w'$ is rejected.
Then the steps (ii)-(v) are repeated until the convergence
(for given $U$ and $E_f$ ) is reached.

Of course, one can move instead of one electron (in step (iv))
two or more electrons,  the convergence of the method can thereby
be improved. Indeed, tests that we have performed for a wide range 
of the model parameters showed that the latter implementation 
of the method, in which $1\le p \le p_{max}$ electrons ($p$ should 
be chosen at random) are moved to new positions, overcomes better the local 
minima of the ground-state energy. This also improves the accuracy of 
the method. Moreover, we have found that the method convergence 
depends sensitively on model parameters and selected values of
$p_{max}$. Thus the first step in our numerical calculations was to 
find the optimum value of $p_{max}$ for the intermediate interactions,
the case studied in this paper.  In particular, we have tested 
the convergence of method for $U = 2$ and four selected values of $p_{max}$,
and namely, $p_{max}=1, 2, 3$ and $N_f$. The results of numerical 
calculations are summarized in Fig.~1, where the quantity 
$M_{0}$ (the minimal number of iterations for which the exact
ground state for given $L$ is reached) is plotted as a function of $1/L$. 
It is seen, that selected $p_{max}$ strongly influence the method
convergence, especially for large $L$, where the cases $p_{max} = 1$ and 
$p_{max} = N_f$ need several times larger number of iterations than the 
cases $p_{max} = 2$ and $p_{max} = 3$. Therefore, we chose the case 
$p_{max} = 2$ for the numerical calculations in the intermediate region.

\section{Results and discussion}
The main aim of this paper is to answer the question, how 
the correlated hopping influences ground-state properties  of the 
one-dimensional FKM in the symmetric case ($E_f = 0, n_f + n_d = 1$). 
In particular, the influence of $t'$ on valence transitions and conducting 
properties of the model was studied. We present results for two selected 
values of the Coulomb repulsion interaction, namely $U=1$ and $U=2$. 
The value $U = 1$ represents the typical behavior of the system for weak 
interactions and the value $U = 2$ represents the behavior 
of the system for intermediate and strong interactions. The 
ground states were studied exactly on the finite clusters consisting of 
$L = 12, 16, 20$ and $24$ lattice sites, while for larger lattices (up to $L = 120$) 
the approximate method, discussed above, was used. 
It is well-known, that the $f$-electron 
density ($n_f$) for the symmetric FKM without correlated hopping is equal to  
$1/2$ and the system is an insulator for all $U > 0$ $\cite {KenLieb}$. It is 
interesting to ask, if the non-zero correlated hopping can change this 
picture.
\vspace*{0.5cm}\\
${\bf U = 1}$
\vspace*{0.5cm}\\
To examine the influence of correlated hopping on the valence transition we
have performed an exhaustive study of the model on finite clusters (up to
$L = 60$) for $t'$ ranging from $-2$ to $2$ with step $0.02$. Results of 
numerical calculations are displayed in Fig.~2 for three selected clusters 
($L = 24, 48$ and $60$) and they clearly demonstrate the strong effects of 
correlated hopping on the valence transitions. For each of these lattices $n_f$ 
exhibits the same characteristic features: (i) On the interval $[-0.38,0.4]$ the 
$f$-electron concentration is equal to $0.5$ (like for the conventional FKM).
(ii) Below $t' = -0.38$ the $f$-electron occupation number increases. 
(iii) Above $t' = 0.4$ the $f$-electron density rapidly falls down 
from $n_f = 0.5$ to $n_{f_{min}} \sim 0.25$ and then gradually increases. 
Due to the fact that we have considered finite (although relatively large) 
lattices the valence transitions have a stair-case structure, which is 
gradually suppressed with increasing $L$, of course with 
the exception of the stair at $t' = 0.4$ that seems to be independent of
$L$.
A detailed analysis showed (see Fig.~2d) that the valence transition from $n_f
= 1/2$ to $n_{f_{min}} \sim 1/4$ is not discontinuous, but it is realized 
through the consecutive intermediate-valence transitions (on the interval
$[0.4,0.44]$) followed by continuous changes of $n_f$  ($L \rightarrow
\infty$) for $t' > 0.44$. The similar behavior was observed
also for negative $t'$ ($t' < -0.38$). 

Moreover, our numerical results showed that there 
are three critical values of $t'$ where $n_f$ changes its character. 
For negative values of correlated hopping $t'_{c_1} \sim -0.38$
was observed. For $t' > 0$ two 
critical values of correlated hopping $t'_{c_2} \sim 0.4$ and 
$t'_{c_3} \sim 0.54$ were found. The numerical results showed that the ground 
state between $t'_{c_1}$ and $t'_{c_2}$ is the alternating configuration. For  
$t' > t'_{c_3}$ the ground states are the segregated configurations with 
different $N_f$. Between $t'_{c_2}$ and $t'_{c_3}$ a few configuration 
types are stable:
\begin{itemize}
\item[(i)] The most homogeneous configurations with $N_f < L/2$.
\item[(ii)] The mixtures of the
alternating configuration ($w_a$) and the empty configuration ($w_0$).
\item[(iii)] The mixtures of the empty configuration ($w_0$) and some 
periodic configurations ($w_p$)
(e.g. $w_{(L=20)} = \{11101110000000000000\}$).
\item[(iv)] Around the boundary of two configuration types $w_A$, $w_B$
mixtures $w_A \& w_B$ are observed.
\end{itemize}
The ground-state configurations  for $t' < t'_{c_1}$ but near $t'_{c_1}$ are 
the most homogeneous configurations with
$N_f > L/2$ and
for smaller values of $t'$ there are the mixtures of the alternating
configuration ($w_a$) and the fully occupied configuration ($w_1$).
Lastly, mixtures of periodic configurations and the alternating configuration 
($w_p \& w_a$) are observed.

Since the ground states of the FKM with correlated hopping consist of 
configurations which are
different from the alternating configuration (that is the ground state 
for the conventional FKM)
it is natural to expect that the correlated hopping will change also the conducting
properties of the model. To verify this conjecture the energy gap ($\Delta =
\lambda_{L-N_f+1} - \lambda_{L-N_f}$) at the Fermi level was calculated
for different finite clusters.
Fig.~3 shows the energy gap on the interval $t' = [-2,2]$ for
two clusters of $L = 36$ and $48$ sites at $U = 1$.
In the region where the  alternating phase  is the ground state 
the gap $\Delta$ has a finite value ($\Delta = U$) independent of $L$, 
and so the system is an insulator. This conclusion is identical with the 
conventional FKM 
($t' = 0$). At $t' \sim 0.4$ the energy gap changes from 
$\Delta = U$ to $\Delta \sim U/2$ for both examined lattices (the inset in
Fig.~3). The detailed analysis showed that the
most homogeneous configurations are the ground states for $t'$ between $t'
\sim 0.4$ and $t' \sim 0.44$ where $\Delta \sim U/2$. 
The extrapolation of the finite-cluster results to the 
thermodynamic limit at $t' = 0.42$ confirmed, 
that the energy gap has a finite value for $L \rightarrow \infty$ and the 
system is really an insulator in this region (Fig.~4). On Fig.~4 we plotted 
also extrapolated results for three other positive values of $t'$ ($t' = 0.5, 
0.55$ and $2$).
Their fits obviously converge to zero (the metallic state) indicating 
that the correlated hopping induces the insulator-metal transition 
at $t' \sim 0.44$. To find the complete picture of metal-insulator transitions 
in the FKM with correlated hopping we have performed also an exhaustive study 
of the model for a wide region of negative values of $t'$ 
($t' \in [-2,-0.38]$). A similar behavior of the model has
been observed. For $t'$ between  $t' \sim -0.5$ and $t' \sim -0.38$, where 
the ground states are the most homogeneous configurations, the energy gap 
has a finite value $\Delta \sim U/2$, while for $t'$ between $t' \sim -1$ 
and $t' \sim -0.5$ the energy gap vanishes in the thermodynamic limit.
This is illustrated in Fig.~5, where the energy gap $\Delta$ is plotted for 
several values of $t'$ as a function of $1/L$. The comprehensive picture of 
insulator-metal transitions in the thermodynamic limit obtained from 
extrapolated behaviors is displayed in Fig.~6 and clearly demonstrates that 
the correlated hopping dramatically changes the conducting properties of the 
model.
In addition to an insulating phase, which characterizes the conventional FKM, 
the large metallic phases are found for the FKM with correlated hopping. 
The metallic phases are observed for positive and also for negative $t'$. For
$t' > 0$ there exists the wide metallic region above  $t' \sim 0.44$ and for 
$t' < 0$ there exists the wide metallic region, between $t' \sim -1$ and 
$t' \sim -0.5$. For positive as well as negative $t'$ the transition from the 
insulating to metallic phase realizes through two discontinuous transitions. 
The first is the insulator-insulator transition from $\Delta = U$ to 
$\Delta \sim U/2$ and the second is the insulator-metal transition from  
$\Delta \sim U/2$ to $\Delta = 0$.  Unfortunately, we were not able to
determine the type of transition (continuous or discontinuous) at $t' \sim
-1$. The finite-size effects near this transition point are still large, and
so it was very difficult to do definite conclusions concerning the
transition type. Our results only indicate that the transition at 
$t' \sim -1$ is probably continuous. 
\vspace*{0.5cm}\\
${\bf U = 2}$
\vspace*{0.5cm}\\
The same procedure as for $U = 1$ has been used also for $U = 2$. As was
mentioned above this value represents the typical behavior of the system
for intermediate and strong interactions. Again we have performed a study 
of the model on finite clusters (up to L = 60) for $t'$
running from $-2$ to $2$ with step  $0.02$. Fig.~7a shows the valence
transition for the cluster with $L = 48$ sites, which qualitatively describes 
$n_f$ for this interaction case. It is seen, that the valence transition has a
similar character as for $U = 1$. In particular, there exists the region 
(between $t'_{c_1} \sim -0.78$ and $t'_{c_2} \sim 0.66$), where the $f$-occupation 
number is constant and equal to $1/2$.
Below $t'_{c_1}$ the $f$-electron occupation number increases from $n_f =
0.5$, and above $t'_{c_2}$ the $f$-electron density rapidly falls down from
$n_f = 0.5$ to $n_{f_{min}} \sim 0.42$ and then gradually increases, similarly as for
$U = 1$. However, there is one important difference. The detailed analysis
performed on finite clusters up to $L = 60$ showed, that the valence transition
at $t'_{c_2}$ changes discontinuously from $n_f = 0.5$ to $n_{f_{min}} \sim
0.42$. Moreover, comparing the valence transitions for 
$U = 1$ and $U = 2$ one can see, that the stronger Coulomb interaction needs 
the larger $t'$ (in the absolute value) to change the ground state from 
$n_f = 1/2$ to another $n_f$. 

Also for this case, we have tried to describe the ground-state configurations.
Although the phase diagram is very complex we were able to describe the basic
configuration types. Between $t'_{c_1}$ and $t'_{c_2}$ the alternating
configuration is the ground state for all investigated lattices. For $t' > 
t'_{c_2}$ the ground states are the segregated configurations with different 
$N_f$. Below $t'_{c_1}$ we were able to specify several configuration types 
with the significant stability regions. In particular:
\begin{itemize}
\item[(i)] The most homogeneous configurations with different $N_f > L/2$.
\item[(ii)] The mixtures of the alternating configuration ($w_a$) and
the fully occupied configurations ($w_1$) of different length.
\item[(iii)] The mixtures of the alternating configuration ($w_a$) and periodic 
configurations ($w_p$).
\end{itemize}

To calculate the energy gap for $U = 2$ the same procedure, the
extrapolation of finite-cluster calculations (up to $L = 120$), was used. The
extrapolated values of the energy gap are plotted in Fig.~7b as a function of
$t'$.
Comparing the energy gaps for $U = 1$ and $U = 2$ (Fig.~6 and Fig.~7b), 
one can see that they exhibit qualitatively same behaviors. In particular, 
there is the wide insulating region around $t' = 0$, two wide metallic regions 
(one for positive and one for negative $t'$) and the insulating region 
below the metallic one for negative $t'$. On the other hand, 
the careful analysis showed one important difference between the case $U=1$
and $U=2$. Namely, for $U = 2$ and $t' > 0$  the system undergoes 
discontinuous insulator-metal transition from $\Delta = U$ to 
$\Delta = 0$, while for $U = 1$ the system undergoes two discontinuous 
transitions (the insulator-insulator transition from $\Delta = U$ to
$\Delta \sim U/2$ and the insulator-metal transition from 
$\Delta \sim U/2$  to $\Delta = 0$). Moreover, one can see that with     
increasing $U$ the metallic regions shift to larger values (in the absolute 
value) of correlated hopping. 

Thus we can conclude that the correlated hopping $t'$ plays the crucial role
in description of the ground-state properties of the spinless FKM in 1D. The
non-zero values of $t'$ substantially change the valence transitions as well
as the conducting properties of the model (in comparison with the conventional
FKM). For the valence transitions it was found that
the relatively small values of $t'$ can induce the large changes of $n_f$
and these changes can be continuous as well as discontinuous.
However, the most important result is that the correlated hopping can
induce the insulator-metal transitions in the FKM for both positive and
negative values of $t'$. These results clearly show that the correlated hopping
strongly influences the ground-state properties of the FKM and so it should not
be neglected in the correct description of real materials with correlated
electrons. Of course, to do a direct comparison with experimental measurements
one has to study FKM with correlated hopping in large dimensions. Work on
this subject is currently in progress.
\vspace*{0.35cm}\\
{\bf Acknowledgments}
\vspace*{0.5cm}\\
This work was partially supported by the Slovak Grant Agency VEGA under Grant
No. 2/3211/23  (P.F.) and the Science and Technology Assistance Agency under 
Grant APVT-20-021602 (H.C.). Numerical results were obtained using computational 
resources of the Computing Centre of the Slovak Academy of Sciences.

\newpage
\begin{center}
{\bf Figure Captions}
\end{center}

\begin{figure}[htb]
\hspace{-2cm}
\includegraphics[angle=-90,width=18.0cm,scale=1]{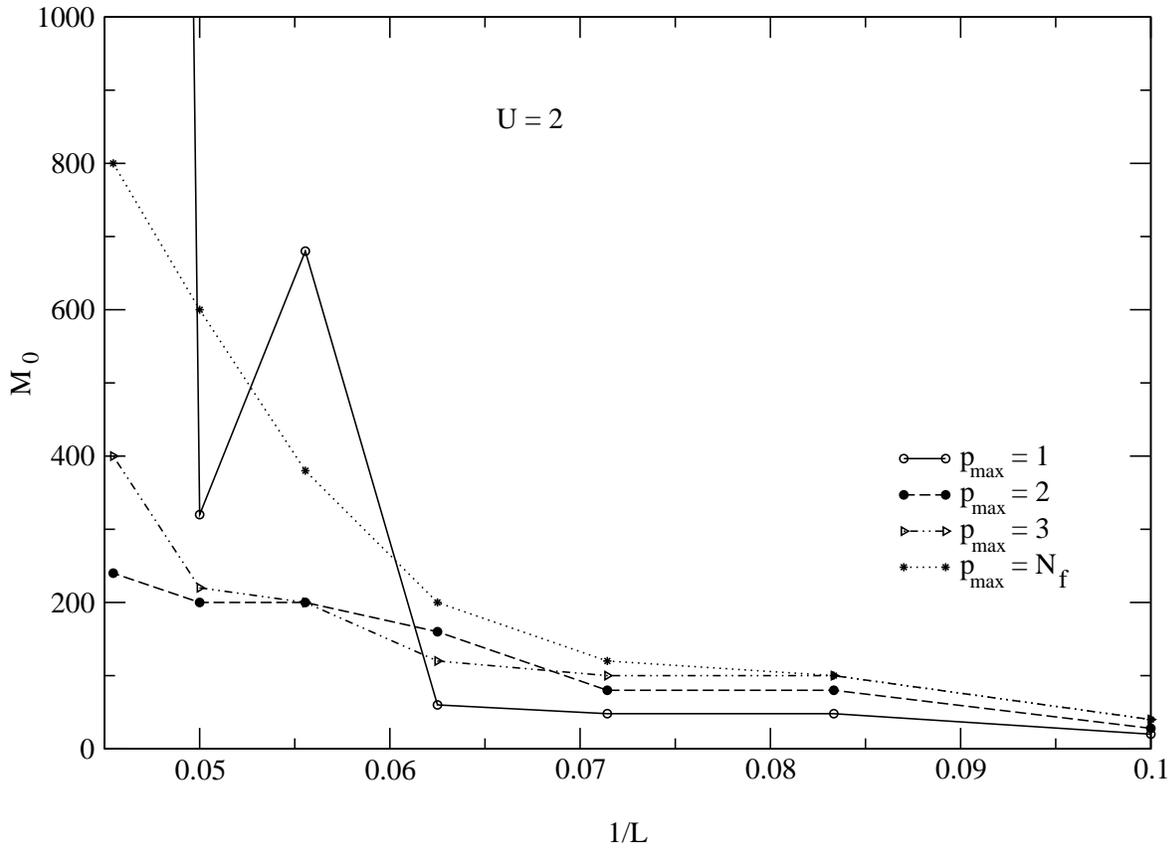}
\vspace{-0.8cm}
\caption{ Dependence of the iteration number $M_{0}$ on $1/L$
for the intermediate Coulomb interaction $U = 2$ and four different values
of $p_{max}$.}
\label{fig1}
\end{figure}

\newpage
\begin{figure}[htb]
\hspace{-2cm}
\includegraphics[angle=-90,width=18.0cm,scale=1]{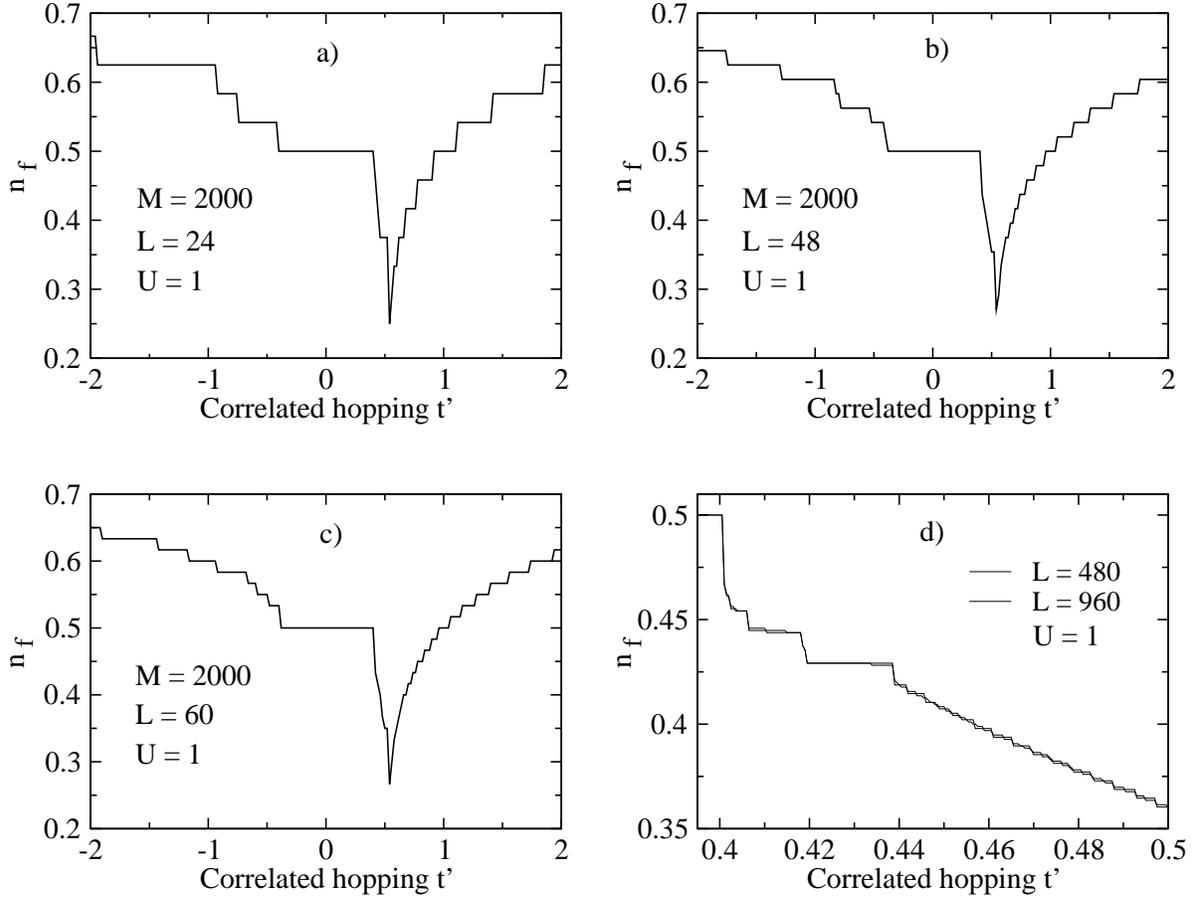}
\vspace{-0.8cm}
\caption{ Dependence of the valence transition on the correlated hopping $t'$ 
for different finite clusters at $U = 1$. For $L = 24$ $(a)$, $L = 48$ $(b)$ 
and $L = 60$ $(c)$ calculations have been done with step $0.02$ and $M = 2000$. 
Results for $L = 480$ and $960$ (d) have been obtained on the extrapolated set 
of ground-state configurations in this region  (the most homogeneous
configurations and the mixtures of the alternating and empty
configurations). }
\label{fig2}
\end{figure}

\newpage
\begin{figure}[htb]
\hspace{-2cm}
\includegraphics[angle=-90,width=18.0cm,scale=1]{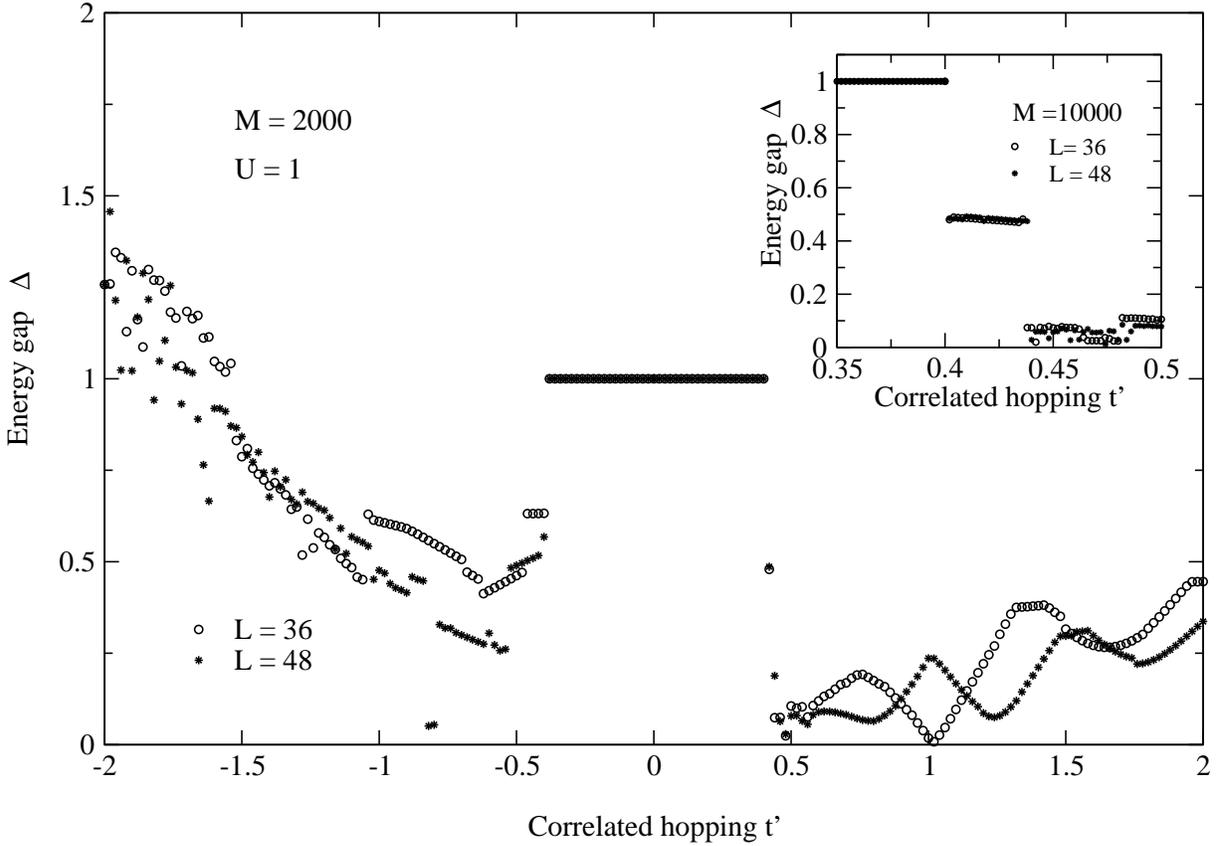}
\vspace{-0.8cm}
\caption{ Dependence of the energy gap $\Delta$ on the correlated hopping $t'$
(calculated with step $0.02$) for two finite clusters of $L = 36$ and $L =
48$ sites ( $U = 1$, $M = 2000$). The inset shows the energy gap 
for the same clusters on the interval $[0.35,0.5]$ calculated 
with step $0.002$ and $M = 10000$. }
\label{fig3}
\end{figure}

\newpage
\begin{figure}[htb]
\hspace{-2cm}
\includegraphics[angle=-90,width=18.0cm,scale=1]{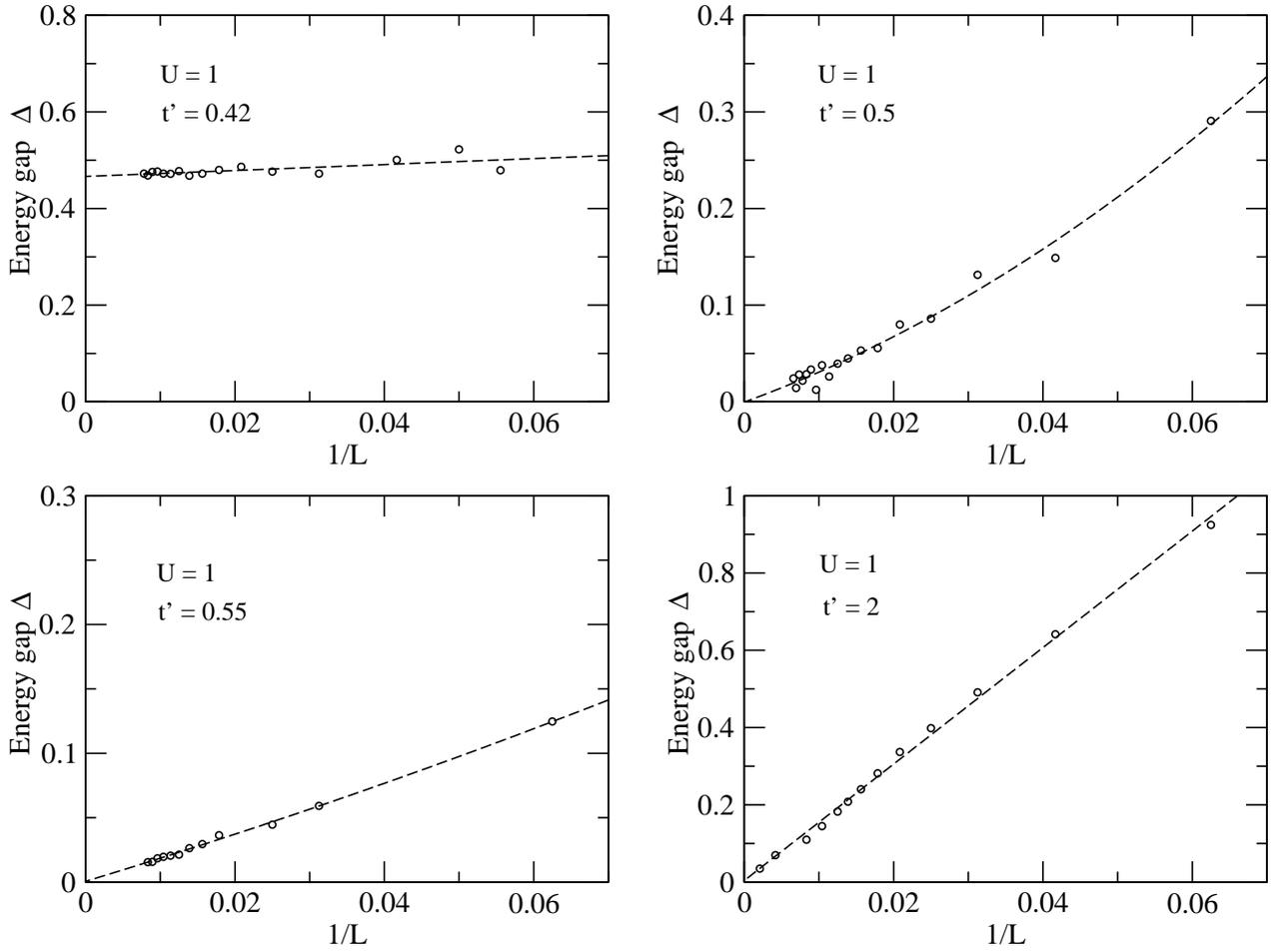}
\vspace{-0.8cm}
\caption{ The polynomial fits of energy gaps for $U = 1$ and four positive 
values of $t'$.  }
\label{fig4}
\end{figure}

\newpage
\begin{figure}[htb]
\hspace{-2cm}
\includegraphics[angle=-90,width=18.0cm,scale=1]{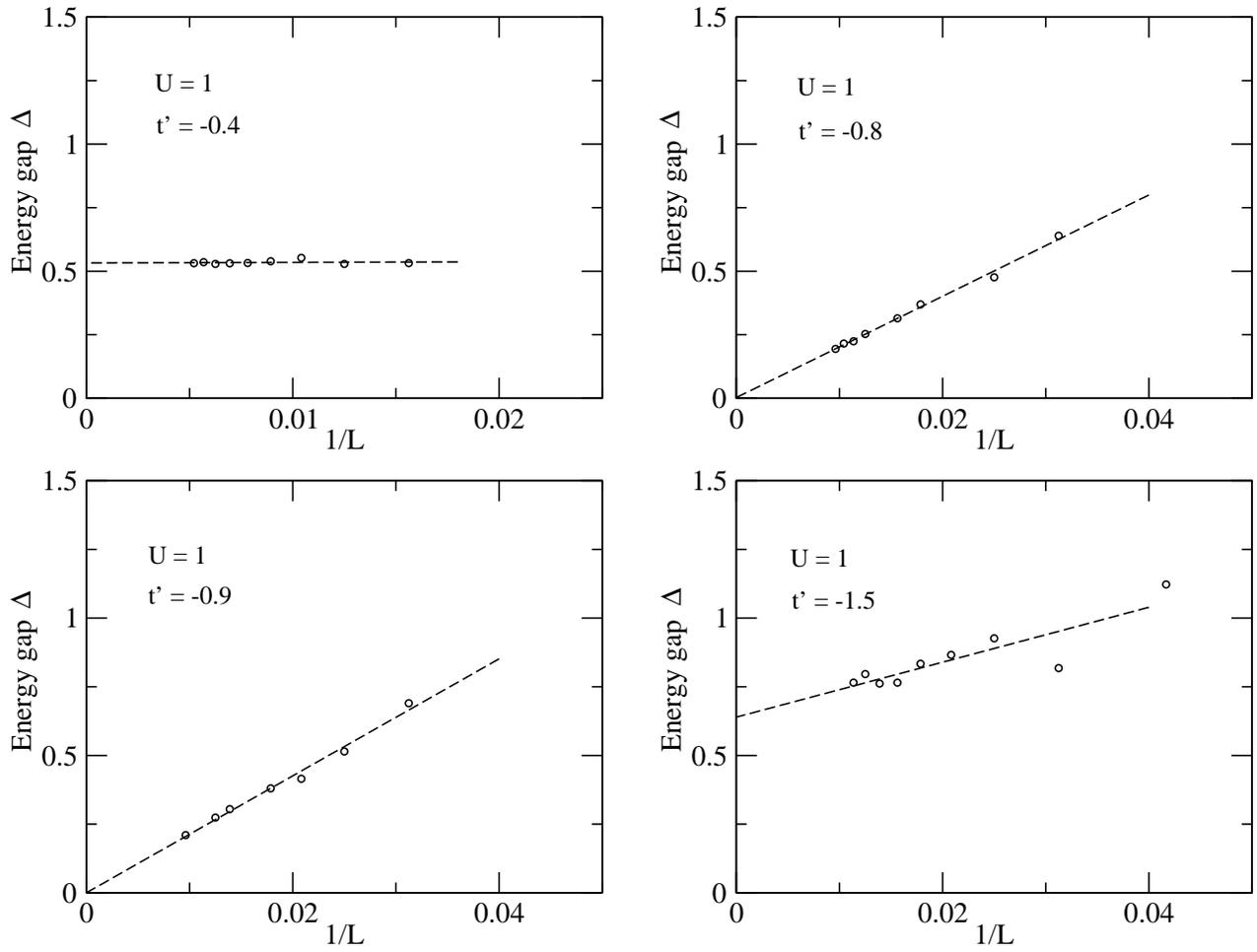}
\vspace{-0.8cm}
\caption{ The polynomial fits of energy gaps for $U = 1$ and four negative 
values of~$t'$.  }
\label{fig5}
\end{figure}

\newpage
\begin{figure}[htb]
\hspace{-2cm}
\includegraphics[angle=-90,width=18.0cm,scale=1]{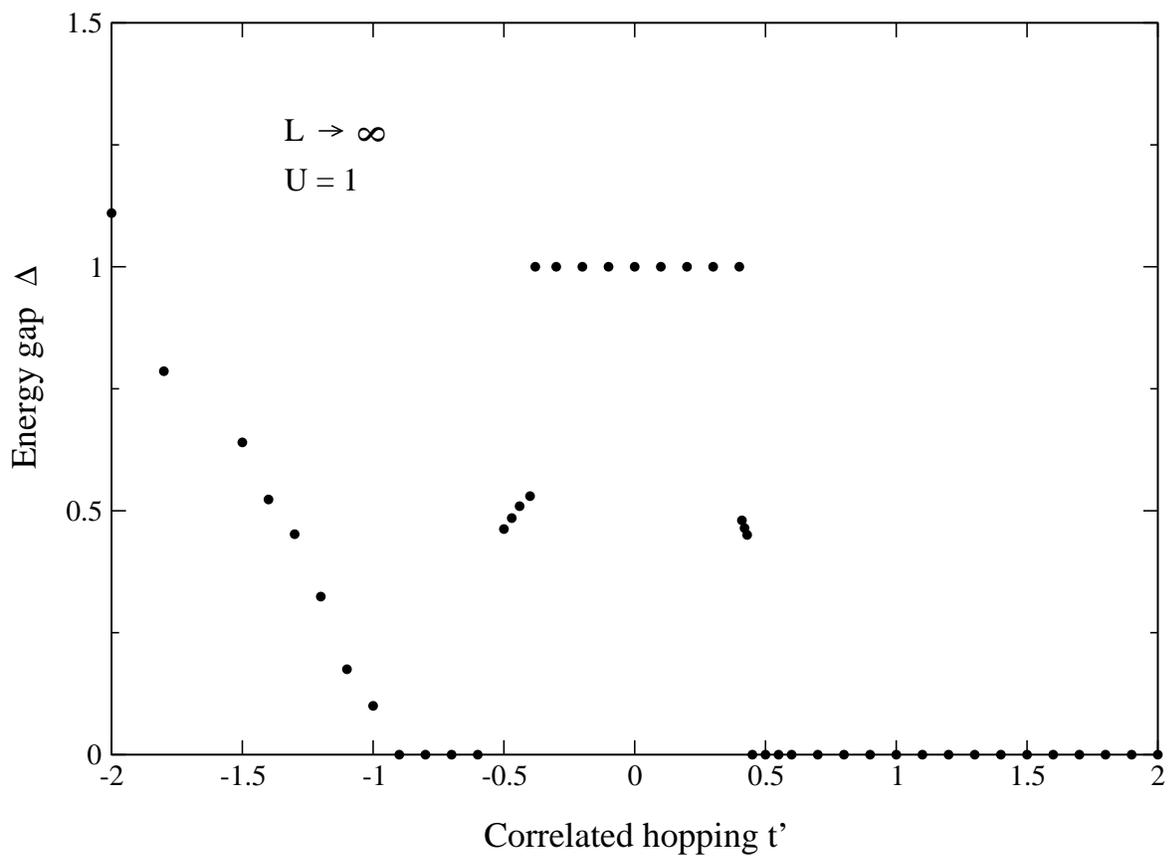}
\vspace{-0.8cm}
\caption{ Dependence of the energy gap $\Delta$ on the correlated hopping $t'$ 
for $L \rightarrow \infty$ and $U = 1$, calculated from extrapolated behaviors.
The data between $t' \sim -0.5$ and $t' \sim 0.44$ were obtained on the extrapolated 
set of ground-state configurations (the most homogeneous configurations). }
\label{fig6}
\end{figure}

\newpage
\begin{figure}[htb]
\vspace{-2cm}
\includegraphics[angle=0,width=16.0cm,scale=0.8]{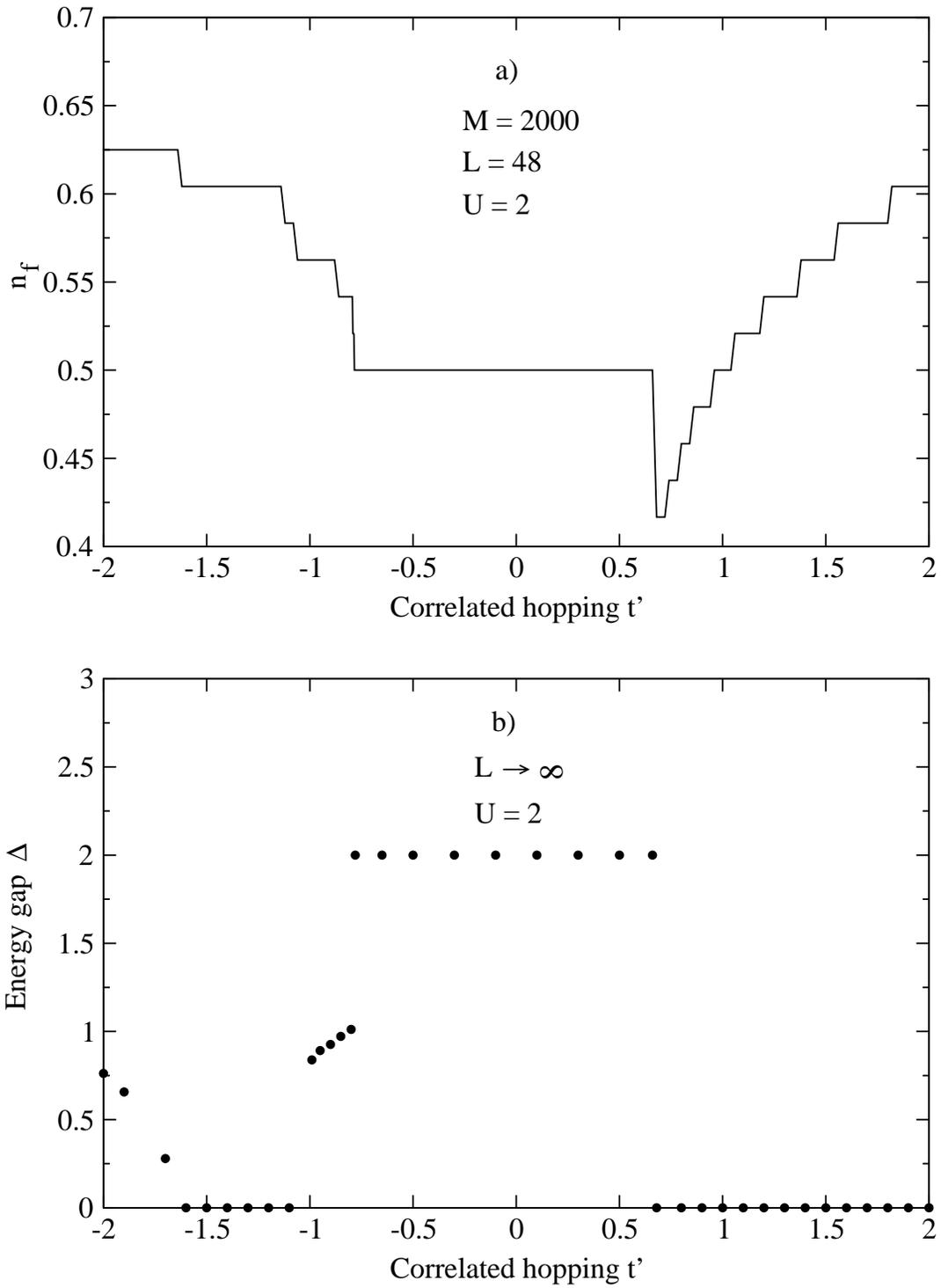}
\vspace{-0.8cm}
\caption{ (a) Dependence of the valence transition on the correlated hopping
$t'$ for the finite cluster of $L = 48$ sites and $U = 2$, calculated with 
$M = 2000$ iterations. 
(b)~Dependence of the energy gap $\Delta$ on the correlated hopping $t'$ for $L 
\rightarrow \infty$ and $U = 2$, obtained from extrapolated behaviors.
The data between $t' \sim -1$ and $t' \sim -0.78$ were extrapolated from the most
homogeneous configurations.   }
\label{fig7}
\end{figure}

\end{document}